\documentclass[12pt,preprint]{aastex}

\newcommand{\ngcseven}{NGC\,752}
\newcommand{\ngcthree}{NGC\,3680}
\newcommand{\teff}{$T_{\mathrm{eff}}$}

\begin{document}

\title{A Spectroscopic Analysis of the Eclipsing Short-Period Binary V505 Persei and the Origin of the Lithium Dip}

\author{
Patrick Baugh\altaffilmark{1}, Jeremy R. King\altaffilmark{1}, Constantine P.
Deliyannis\altaffilmark{2,4}, and Ann Merchant Boesgaard\altaffilmark{3,4}
        }

\affil{
\altaffiltext{1}{Department of Physics and Astronomy, 118 Kinard Lab, Clemson 
University, Clemson, SC 29634-0978; pbaugh@clemson.edu, jking2@clemson.edu}
\altaffiltext{2}{Department of Astronomy, Indiana University, Swain Hall West 
319, 727 East 3rd Street, Bloomington, IN 47405-7105  USA; 
con@astro.indiana.edu}
\altaffiltext{3}{Institute for Astronomy, University of Hawaii at Manoa, 2680 
Woodlawn Drive, Honolulu, HI 96822, USA;  boes@ifa.hawaii.edu}
\altaffiltext{4}{Visiting Astronomer, W.M KECK Observatory, which is operated 
as a scientific partnership among the California Institute of Technology, the 
University of California and the National Aeronautics and Space 
Administration.  The Observatory was made possible by the generous financial 
support of the W.M. Keck Foundation.}
     }
     
\begin{abstract}
As a test of rotationally-induced mixing causing the well-known Li dip in
older mid-F dwarfs in the local Galactic disk, we utilize high-resolution and -S/N Keck/HIRES
spectroscopy to measure the Li abundance in the components of the
${\sim}1$ Gyr, [Fe/H]$=-0.15$ eclipsing short-period binary V505 Per.  We
find $A(\mathrm{Li})\sim2.7\pm0.1$ and $2.4\pm0.2$ in the
\teff${\sim}6500$ and 6450K primary and secondary components, 
respectively.  Previous $T_{\rm eff}$ determinations and uncertainties
suggest that each component is located in the midst of the Li dip.  If so, 
their A(Li) are ${\ge}$2--5 times larger than A(Li) detections 
and upper limits observed in the similar metallicity and
intermediate-age open clusters \ngcseven\ and 3680, as well as the more
metal-rich and younger Hyades and Praesepe. These differences are even
larger if the consistent estimates of the scaling of
initial Li with metallicity inferred from nearby disk stars, open clusters, and recent
Galactic chemical evolution models are correct.  Our results suggest, 
independently of complementary evidence based on Li/Be ratios, Be/B ratios, 
and Li in subgiants evolving out of the Li dip, that main-sequence angular 
momentum evolution is the origin of the Li dip.  Specifically, our stars' 
A(Li) indicates tidal synchronization can be sufficiently efficient and occur 
early enough in short-period binary mid-F stars to reduce the effects of 
rotationally-induced mixing and destruction of Li occuring during the main-sequence 
in otherwise similar stars that are not short-period tidally-locked binaries.
\end{abstract}


\section{INTRODUCTION} 
Lithium is of fundamental astrophysical importance.  This fragile 
light element is quickly destroyed in stars when 
exposed to $T{\ge}2.5{\times}10^6$ K, making it useful in studying 
matter transport in stars and providing observational feedback on our 
understanding of stellar structure and evolution 
\citep[for a review, see][]{pinsonneaultarap}.  In addition, stellar Li 
abundances have cosmological applications.  Combining the 
A(Li)\footnote{${\rm A}(Li){\equiv}{\log}N({\rm Li})$ on the usual scale
where the logarithmic number abundance of hydrogen is given by 
${\log}{N}({\rm H}){\equiv}12$.} 
in 
old stars with a) accurate stellar evolution models to account for stellar 
destruction, and b) accurate chemical evolution models to account for Galactic 
production, should lead to the primordial A(Li), which provides 
constraints on Big Bang nucleosynthesis \citep{BS85} and an independent check 
on cosmological parameters determined by WMAP \citep{WMAPp}.  However, 
the vulnerability of Li to destruction that makes it a good probe of stellar 
interiors also complicates the derivation of {\it initial} halo 
dwarf A(Li)--our best connection to the Big Bang A(Li).

The astrophysical usefulness of Li, therefore, strongly depends on having 
correct stellar models that accurately trace the {\it in situ} history of Li.  
Standard\footnote{That is, spherically symmetric models that ignore 
rotation, diffusion, mass loss, magnetic fields, and other physics 
that could affect real stars \citep{DDK90,pinsonneaultarap}.} 
stellar models (SSMs) suggest that, for stars of about 1.2$M_\odot$, 
the observed surface A(Li) can go down only, 
a) during the early pre-main sequence (PMS) when 
some of the interior Li is actually destroyed, and b) during subgiant 
evolution as the deepening surface convection zone (SCZ) dilutes it.  Standard 
theory predicts no surface Li depletion during the main sequence (MS).  
More specifically, since Li burns at relatively low temperatures, 
anytime the SCZ reaches deep enough into the star's 
interior, visible Li depletion will occur at the surface.  This occurs during 
the PMS phase and, for stars of around 1.2$M_\odot$, the SCZ is not deep 
enough during the MS to carry Li into the interior regions where it can be 
burned appreciably.  Even during late PMS convective burning, such depletion 
is minimal.  For a 1.2$M_\odot$ star with [Fe/H]${\le}-0.1$, the 
maximum expected depletion on both the MS and the PMS is $<0.1$ dex 
\citep{pinsonneaultarap}.  Therefore, little Li depletion in stars of this 
mass (or higher) is expected on the PMS, and virtually none on the MS, based 
on the predictions from SSMs \citep{conferencep}. 

In contradistinction, striking Li \citep[and Be;]{BK02} depletion is observed 
in disk MS stars with $6300{\leq}T_{\mathrm{eff}}\leq6900$ K.  In the Hyades, \citet{boesgaardp} 
discovered that stars with $6500{\leq}T_{\mathrm{eff}}\leq6850$ K were 
depleted by ${\geq}0.5-2$ dex as compared to stars merely 200--300 K hotter 
or cooler.  This abrupt F-star Li(Be) depletion (the Li dip) occurs mainly 
{\it during the MS}, since much younger clusters such as the Pleiades 
(${\sim}100$ Myr) show a nearly flat Li-\teff\ relation \citep{BBR88,DRS93}.  
The Li dip presents a challenge to SSMs, which predict negligible 
Li depletion on the MS, and casts doubt on their ability to 
either predict or backtrack Li abundances to initial values. Instead, 
stellar models are required that can explain the Li dip 
accurately by modeling Li abundances over a star's lifetime.  
A number of refinements to SSMs have been proposed to explain the Li dip 
\citep{balachandranp,Deli2000,SR2005}, but here we focus on 
rotationally-induced mixing (RIM) \citep{RIMp,RIM2p}.  RIM models propose that 
a loss of angular momentum, as stars spin down on the MS, could cause slow 
mixing between the photosphere and the hotter denser region beneath the SCZ, 
resulting in the depletion of surface Li.

To test this theory, \citet{Litestp} suggest that Li abundances of 
short-period tidally-locked binaries (SPTLBs) be measured.  RIM and stellar tidal 
theory \citep{ZBp} together predict that a SPTLB loses most of its angular momentum 
during the very early PMS, when interior densities and temperatures are too low to 
burn Li.  Hence, lacking significant angular momentum to lose and drive future
requisite mixing, neither does it suffer MS Li depletion responsible for the Li dip. 
Thus, one method to test if angular momentum transfer is indeed the causal force 
behind the Li dip is to find SPTLBs whose ages and temperatures should place them in 
the Li dip, and measure their Li.  A value of A(Li) higher than otherwise similar Li dip 
stars would support RIM models.  V505 Persei, a Pop I intermediate-age short-period binary 
system, is made up of two F stars of nearly equal mass, radius, and \teff\ 
\citep{marschallp, tomasellap}.  Both masses, \teff, and ${\sim}4$ day period 
make them good candidates to test if an early PMS loss of angular momentum can 
later preserve Li compared to otherwise similar non-SPTLB stars.  We measure 
the A(Li)s of V505 Per and compare them in the A(Li)
versus zero-age MS (ZAMS) \teff\ plane with stars of intermediate-age open 
clusters, showing in the process that the components of V505 Per have 
A(Li)s higher than single stars of their ZAMS \teff. 

\section{DATA AND ANALYSIS}
\subsection{Observations}
High-resolution ($R=45,000$) spectroscopy of HD 14384 was obtained on 30 
and 31 August 1997 using Keck/HIRES.  
Three nightly exposures totalled 8 and 5.5 
minutes on the first and second nights, respectively. 
Debiasing, flat-fielding, scattered light removal, order 
tracing/extraction, and wavelength calibration were carried out with standard 
routines in IRAF.

Given the short integration spans relative to the 4.2 d period, 
we co-added the 3 nightly spectra.  
The resulting ${\lambda}6707$ continuum level Poisson SNR/pixel 
is 850 and 580 on the first and second 
nights.  Spectra were fit with a low order polynomial to perform 
a continuum normalization.  Figure 1 shows our co-added, normalized 
spectra.  The physically similar components of the SB2 are 
distinguished and identified both by slight 
differences in the strengths of the ${\lambda}6717$ CaI feature and the 
relative Doppler shifts expected from the orbital ephemeris of 
\citet{tomasellap}.  The relative Doppler shifts of the components in each 
night's coadded spectrum, needed for the spectrum synthesis described next, 
were measured from the centroids of the CaI features.

\subsection{Syntheses and Comparison}
To derive Li and Fe abundances, we conducted spectrum synthesis of the 
${\lambda}$6707 \ion{Li}{1} region to account for the line blending between 
the stellar components.  Such synthesis requires knowledge of \teff, ${\log} g$, 
radius ($R$), and microturbulent velocity ($\xi$) of each star.  The first
of three of these parameters are taken from \citet{tomasellap}, who determine radii 
and associated uncertainties from a \citet{WD71} code-based solution with
updated model stellar atmospheres \citep{MSK92}; these radii agree with 
alternate solutions derived from the radial velocities and $BV$ photometry
alone to within 0.05 $R_{\odot}$, which is well within their stated ${\pm}0.11$ 
$R_{\odot}$ uncertainty adopted here. The \teff values and uncertainties 
of \citet{tomasellap}, $6512{\pm}21$ and $6462{\pm}12$ K for the A and B components, 
are derived from multiparameter ${\chi}^2$ fitting of 700 {\AA} of their own high-resolution 
($R{\sim}$20,000) spectra against synthetic spectra grids.  Encouragingly, the \teff {\ }differences 
of the components they derive as part of the orbital solutions agree to within 18 K of the 
spectroscopically-derived value.  

The \citet{tomasellap} masses and radii yield  ${\log} g=4.33{\pm}0.02$ for both components.
We determined $\xi$ (1.73 and 1.70 km s$^{-1}$) using the \teff- and ${\log}g$-dependent calibration 
of  \citet{microturbulencep}.  We note, though, the derived Li abundances are insensitive to
the value of log $g$ and $\xi$.  .  While [Fe/H] was eventually determined from a comparison of 
the observed and synthetic spectra, we adopted an initial metallicity of [Fe/H]$=-0.35$ from the 
photometric determination of \citet{nordstromp}.  

We used the \teff, log g, and [Fe/H] values to 
interpolate model atmospheres from the grids of \citet{kuruczp}.  MOOG\footnote{http://as.utexas.edu/~chris/moog.html} 
was used to create a synthetic spectrum for each star using the ${\lambda}6707$ linelist of 
\citet{kingp}.  The spectra were smoothed by convolving with a Gaussian with FWHM  measured 
from clean, weak lines in the observed spectra.  Each component's synthetic spectrum was Doppler 
shifted and then combined, using the product of the square radii and the Planck function value at 
6707{\AA} (a pseudo-monochromatic luminosity) as a weighting factor. This correction for flux dilution
is appropriate for the first night's observations (orbital phase of 0.74) since each star's 
undiminished flux contributes to the spectrum.  The orbital phase (0.97) of second night's observations 
places the system on the cusp of primary eclipse; however, the eclipses are very sharp, and the 
total system flux diminishment in $B$ at this phase is $\lesssim$0.01 mag of the ${\sim}0.5$ mag total at primary eclipse. 

Finally, we compared the synthetic spectra to the observed spectra using $\chi^2$ minimization methods.
We used the ${\lambda}6705$ \ion{Fe}{1} line to determine [Fe/H], forcing both 
stars to have an assumed identical Fe abundance.  Once a best-fit value of [Fe/H] 
had been determined, we moved on to A(Li), which was allowed to differ in 
the two components to allow for possible differing Li depletion in the two stars.  
These synthetic spectra are compared to the observed data in Figure 1.

\subsection{Results}
The analysis of the first night's data yielded [Fe/H]$=-0.15{\pm}0.03$.  The 
quoted error is due to the 1${\sigma}$ level fitting uncertainties alone; even 
so, our metallicity estimate is in good agreement with 
[M/H]$=-0.12{\pm}0.03$ from \citet{tomasellap}.  The metallicity from 
the second night was not calculated due to the unfortunate placement of the 
${\lambda}6705$ FeI feature in the secondary star with a 
detector/reduction artifact that can be seen in Figure 1; in carrying out the 
Li syntheses, we assumed the [Fe/H] value from the first night's data.  

The Li syntheses (Figure 1) yield average A(Li) of 
$2.67{\pm}0.1$ and $2.42{\pm}0.2$ for the primary and secondary components, 
respectively.  While the best-fit Li abundances differ by only a few hundredths 
of a dex, the larger quoted uncertainties in A(Li) are 
dominated by uncertainties in the continuum location. Contributions from 
fitting uncertainties in the $\chi^2$ minimization and \teff\ uncertainties 
amount to only ${\pm}0.03-0.06$ dex and ${\pm}0.01-0.02$ dex, respectively; 
contributions from uncertainties in ${\xi}$ and log $g$ are similarly small 
or smaller and are ignored.   Abundance uncertainties arising from uncertainties 
in the flux dilution factors of each component, which in turn arise from uncertainties 
in \teff\ and the \citet{tomasellap} stellar radii, are $0.01-0.02$ dex.

\section{DISCUSSION}

\subsection{Li versus ZAMS $T_{\rm eff}$}
Interpreting the A(Li) of our SPTLB components requires that 
they be placed in the context of the Li dip morphology defined by other single 
(or non-SPTLB) stars.  In her comparisons of the Li dip in various open 
clusters, \citet{balachandranp} found that the \teff\ at which the Li dip 
occurs varies based on metallicity, but that the ZAMS \teff\ at which the Li 
dip is located does not; this provides a means by which the Li dip 
morphology of different populations of disk stars can be compared.  
Additionally, \citet{balachandranp} found that the morphology of the cool side 
of the Li dip is age-dependent.  Some evidence suggests that the Li dip may 
begin to form as early as an age of 150 Myr \citep{SD2004} or even 100 Myr 
\citep{Marg2007}.  Clearly, comparing our stars with bona fide Li dip stars 
requires knowledge of our SPTLB components' ZAMS \teff\ and age. 

For consistency, we followed the approach of \citet{balachandranp} to find the 
ZAMS \teff\ of each star by looking at differences implied by isochrones (and 
their assumed color transformations) between our stars in their current 
evolutionary state and on the ZAMS.  This required that we first determine the 
age of our SPTLB stars, which we did by placing the components in the radius 
versus mass plane and comparing these locations with sequences from 
[m/H]$=-0.14$  Yonsei-Yale isochrones \citep{Y2isop}.  As shown in Figure 2, 
this implies an age of $1.15{\pm}0.15$ Gyr for the system; the majority of the 
age uncertainty comes from uncertainty in the radii estimates.  

We then used the Yonsei-Yale isochrones with the Green color-temperature relations, 
which are also employed by the Revised Yale Isochrones \citep{isochronep} used 
by \citet{balachandranp}, to determine the difference 
between the \teff\ at 1.15 Gyr and on the ZAMS. This \teff\ 
difference was then applied to our current \teff\ value from \citet{tomasellap}, yielding 
ZAMS \teff\ values of 6483 and 6432K for the primary and secondary 
components.  The $1{\sigma}$ level uncertainties in our interpolation of the 
\teff-mass relations of the isochrones are ${\le}12$ K.

Figure 3 presents the Li-ZAMS \teff\ diagram containing: {\ }the V505 Per 
components; the literature data reanalyzed by \citet{balachandranp} for the 
open cluster \ngcseven, having a $1.45$ Gyr age \citep{AT2009} and 
[Fe/H]$=-0.15$ \citep{NGC752p}; the Li data for the 1.75 Gyr, 
[Fe/H]$=-0.08$ cluster \ngcthree\ \citep{AT2009}; and the Li data of 
\citet{balachandranp} for the 650 Myr [Fe/H]$=+0.13$ Hyades cluster.  We 
determined the \ngcthree\ object masses using a Legendre polynomial relation 
to map the dual $V$ magnitude and mass abscissas of Figure 4 of \cite{AT2009}.  
We then used the Yonsei-Yale isochrones as described above to determine the \teff\ 
difference between the ZAMS and at 1.75 Gyr at a given mass.  This difference 
was then applied to the {AT2009} \teff\ values to yield ZAMS \teff\ values.  

\subsection{v505 Per versus the Hyades}

If the v505 Per $T_{\rm eff}$ values and uncertainties of \citet{tomasellap} are reliable, 
and if the {\it relative\/} uncertainties in the to-ZAMS $T_{\rm eff}$ corrections for the 
v505 Per components compared to those of the open cluster comparison stars are not 
several times the size of the corrections themselves (${\sim}30$ K for our binary components 
and ${\sim}40$ K for similar mass stars in the slightly older NGC 752 cluster), then 
Figure 3 indicates that both our SPTLB components 
are positioned inside the Li dip that is well defined by the Hyades data \citep[or those of the
or Praesepe data not shown here; see Figure 12 of][]{balachandranp}.  The larger Li abundances 
in the v505 Per components compared to nearly all of the  younger and more metal-rich Hyades 
data in Figure 3 is especially notable given the metallicity difference and the age difference, 
which we discuss in turn, between the two.  

First, comparisons of the v505 Per Li abundances with those in cluster stars are most meaningful 
if some account of initial Li abundance differences can be made. An empirical 
approach to parameterize Galactic disk Li production in terms of Fe production is to
use the upper envelope of the Li-Fe relation exhibited by large samples of field stars. The field 
star data over the range $-1{\leq}\mathrm{[Fe/H}{\leq}0$ in Figure 7 of 
\citet{LHE} suggest an initial Li-to-Fe (logarithmic by number) relation in 
the local disk having slope ${\sim}1$ dex/dex.  For comparison, the Galactic 
chemical evolution model in Figure 9 of \citet{Travagliop} produces a slope of 
${\sim}0.7$ dex/dex over the same [Fe/H] range, though this slope may be too 
small since it is unable to reproduce the initial solar Li abundance.  Indeed, determinations
of the slope of the Li-to-Fe relation using Li abundances on the G-dwarf Li peak of various open 
clusters are significantly larger at 1.4 dex/dex \citep{Boes91} and 1.0 dex/dex \citep{Cumm11}.    

The Hyades and Praesepe have super-solar metallicities:  [Fe/H]${\sim}+0.10$ to $+0.15$ 
\citep{Boes89,BF90}.  The local disk field and \citet{Cumm11} open cluster Li-Fe relation 
thus implies the initial Hyades and Praesepe Li abundances were a factor of 2 larger than 
for the v505 Per components; the \citet{Boes91} open cluster Li-Fe relation implies initial 
Li abundances a factor of 2.6 larger than for the v505 Per components.  Accounting for initial 
Li differences in this way makes  the observed present-day difference between v505 Per and 
Hyades Li abundances even more remarkable.  

Second, the red side of the Li dip is known to flatten with increasing age, which may be due
to increasing Li depletion in the dip stars with age \citep{balachandranp}.  Nevertheless, 
despite their older age, our SPTLB components exhibit Li abundances a factor of 2 {\it larger} 
than nearly all Li detections or upper limits at similar ZAMS {\teff}  in the younger ($\sim$650 Myr) 
Hyades and Praesepe.  Once reaching the v505 Per age, stars on the red side of the dip 
in these clusters would presumably have even lower Li abundance than at present.  

\subsection{Comparison with other data}
  
Figure 3 also compares V505 Per with stars of more similar [Fe/H]: those in \ngcseven\ 
and NGC 3680.  The A(Li) of our SPTLB components are a factor of 2-5 larger than the upper 
limits for the \ngcseven\ and NGC 3680 Li dip stars (one 3680 star on the steep blue side 
of the dip is within the error bar of our primary component).  Note that \cite{SRP94} argue 
that their high-resolution spectroscopy of solar-type dwarfs in \ngcseven\ suggests 
[Fe/H]$=+0.01$ for this cluster, 0.15 dex larger than the canonical value quoted by \citet{NGC752p}.  
If so, the above discussion of the initial Li abudnance versus [Fe/H] relation suggests an even 
larger difference between the Li depletion factors of our SPTLB components and the \ngcseven\ Li 
dip stars.

Finally, it is important to recall that the preservation of Li by SPTLBs 
occurs if angular momentum loss occurs sufficiently early during the PMS, when 
interior temperatures and densities are too low to burn Li.  This early 
synchronization is predicted to occur in systems with periods below some 
critical period that is a function of stellar mass and metallicity 
\citep{Zahn94}.  The models of \citet{Zahn94} for Galactic disk stars of mass 
1.2M$_{\odot}$ indicate this critical period is 6 days.  The 4.2d period of 
V505 Per falls under this critical period, though the 1.25 and 
1.27M$_{\odot}$ components slightly exceed the 1.2M$_{\odot}$ maximum mass 
considered by the modeling of \citet{Zahn94}.  The Hyades SPTLB vB 34 resides 
within the blue region of the Hyades Li gap, but does not clearly demonstrate 
a Li abundance larger than similar single stars \citep[see, e.g., Figure 1 of][]{Litestp}.  
However, Yonsei-Yale isochrones ([Fe/H]$=+0.13$, 650 Myr for the Hyades; 
[Fe/H]$=-0.14$, 1.15Gyr for v505 Per) indicate that the vB 34 masses are 0.20-0.23M$_{\odot}$ 
larger than for v505 Per; thus, the 3.1d period of vB 34 may not be below a necessary 
critical period for its larger mass components. 

\section{CONCLUSIONS}
Lithium is important for probing stellar interiors and evolution.  It can be used to 
study transport of matter in the stars, Galactic chemical evolution, and BBN.  Such uses 
depend heavily on accurate predictions of Li abundance evolution within stars. 
It is known that SSMs are unable to explain the Li dip in disk mid-F dwarfs. 
Modified stellar models that include the action of rotationally-induced slow 
mixing can explain the Li dip as the result of slow Li mixing in stars 
that are currently undergoing angular momentum loss and are sufficiently 
massive that interior temperatures and densities are sufficiently large to 
burn Li as a result of such mixing. 

Although other types of physical mechanisms have also been proposed as 
explanations of the Li dip, including diffusion, mass loss, and other types of 
mixing, a variety of observational evidence favors, often quite strongly, the 
RIM-type models over other mechanisms.  This evidence includes (but is not 
limited to): a) the Li/Be depletion correlation where {\it both} elements are 
depleted, but Li more severely \citep{Stephensp,Deli98,BAKDSp} {\,}b) the 
Be/B depletion correlation \citep{Boes2005} {\,}c) subgiants in M67 revealing 
the size and shape of the (MS) stellar preservation region as they evolve 
out of the cool side of the Li dip \citep{SD2000} {\,}d) and the early 
MS formation of the Li dip.

Here, we present an independent observational test of the RIM explanation of 
the Li dip.  The $T_{\rm eff}$ values and uncertainties of \citet{tomasellap} 
indicate that the components of the mildly metal-poor 
([Fe/H]${\sim}-0.15$) intermediate-age (${\sim}1.1$Gyr) short-period 
tidally-locked binary V505 Per both reside in the Pop\,I Li dip defined by 
open cluster observations (assuming the differential ${\sim}10$ K to-ZAMS
$T_{\rm eff}$ corrections suggested by isochrones for our 
stars relative to the younger Hyades and older NGC 752 and NGC 3680 clusters 
are not, in fact, an order of magnitude larger).  If angular momentum loss in such a system occurred 
very early during the pre-main sequence phase, then it would suffer no or 
reduced RIM during the MS compared to non-SPTLB stars occupying the Li dip 
region; as a result, the V505 Per components would exhibit larger Li 
abundances than otherwise similar stars in the Li gap.  

We find that 
the V505 Per components' Li abundances are at least 2-5 times 
larger than both:  a) the Li upper limits in the ${\sim}1.5-2$Gyr and 
similarly mildly metal-poor ({Fe/H}${\sim}-0.15$ and $-0.08$) clusters 
\ngcseven\ and 3680, and  b) the upper limits and Li detections in the younger 
metal-rich Hyades and Praesepe clusters.  If there exists an initial Li-Fe 
relation of positive slope, as field star data and open cluster observations and Galactic chemical evolution 
models each independently suggest, and initial Li abundances indeed scale with [Fe/H] in the 
recent disk, then the Li overabundance of V505 Per is even more dramatic in 
the case of the younger clusters (which presumably started with a higher 
initial Li abundance) and perhaps in \ngcseven\ if one assumes the higher [Fe/H] 
value of \citet{SRP94}. 

Our results suggest, independently, that angular momentum evolution on the 
MS is responsible for the Li dip, confirming the conclusions drawn from the 
variety of observational evidence listed above, involving both field and 
cluster dwarfs.  SPTLBs with higher-than-normal Li have been found in the 
650-Myr old Hyades \citep[][see Soderblom et al. 1990 for a related 
idea]{THDP93}, the 4.5Gyr-old M67 \citep{Deli94}, and in other contexts 
\citep{Litestp}; and, 
significantly, SPTLBs have {\it normal} Li in clusters, such as the Pleiades, 
which are too young for much RIM-related depletion to have occurred 
\citep{Litestp}.  Our results for V505 Per complement these previous findings 
in that the V505 Per SPTLB stars are the hottest high-Li SPTLBs discovered so 
far, and are indeed very close to the limiting \teff\ beyond which the models 
of \citet{Zahn94} can no longer synchronize both components sufficiently early 
during the pre-MS to prevent Li destruction.  The SCZ of hotter 
SPTLBs is too shallow and the their Hayashi paths too short for the components 
to grab onto each other sufficiently effectively so as to cause 
tidal locking during that same evolutionary phase.   

While the Li abundances in our V505 Per components likely reflect some Li 
depletion from a plausible initial abundance in the range 
$A(Li)=3.0-3.3$, which is not unexpected 
\citep[see section 2.2.1 of][]{Litestp}, they nevertheless suggest that early 
tidal circularization can be efficient in mid-F stars in the Li dip and 
mitigate the effects of RIM on Li depletion.  Identification of additional Li 
dip SPTLBs, analysis of their A(Li), and comparison with that of 
single stars of similar Li dip position, metallicity, and age would be a 
profitable means to extend these conclusions.  

\acknowledgments
This work was supported by NSF awards AST-0239518 and AST-0908342 to JRK, 
and AST-0607567 to CPD. We thank Bruce Twarog for providing us with the 
\ngcthree\ data.  

{\it Facility:} \facility{KECK}

\begin{figure}
\plotone{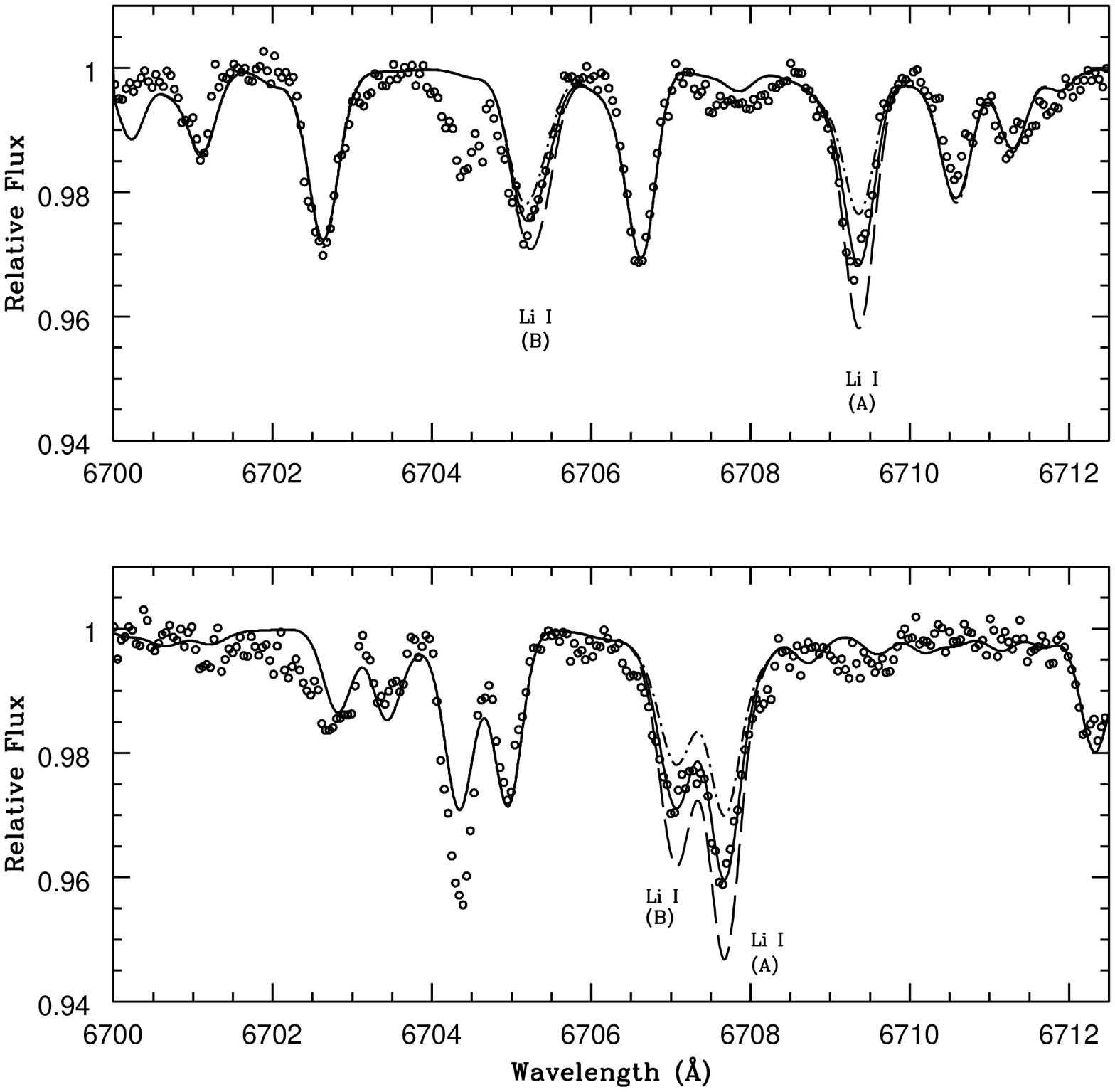}
\vskip -2.0cm
\caption[Li abundance]
{Final synthetic spectrum (solid line) overlaid the observed spectrum (dots), 
with two other synthetic spectra offset in Li abundance by $\pm$0.15 dex for 
comparison (dashed and dotted lines).  The top and bottom panels show the 
coadded spectra for 30 Aug and 31 Aug, respectively.  The ``feature'' at 6704.5{\AA} 
is an artifact that contaminates the \ion{Fe}{1} line of the B 
component in the second night's spectrum.}
\label{fig:fig1}
\end{figure}

\begin{figure}
\plotone{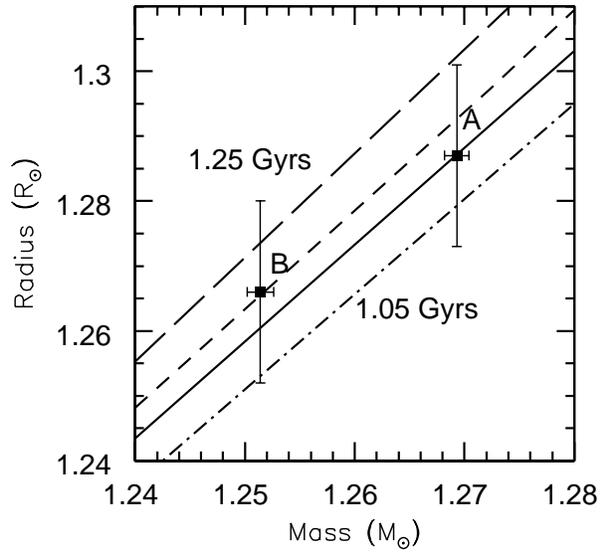}
\vskip -6.0cm
\caption[Isochrones]
{Yonsei-Yale isochrones (lines) for [Fe/H]$=-0.14$ and [${\alpha}$/Fe]$=0$ 
from 1.05-1.25 Gyr plotted in the radius versus mass plane with the physical 
determinations of the V505 Per A and B components.} 
\label{fig:fig2}
\end{figure}

\begin{figure}
\plotone{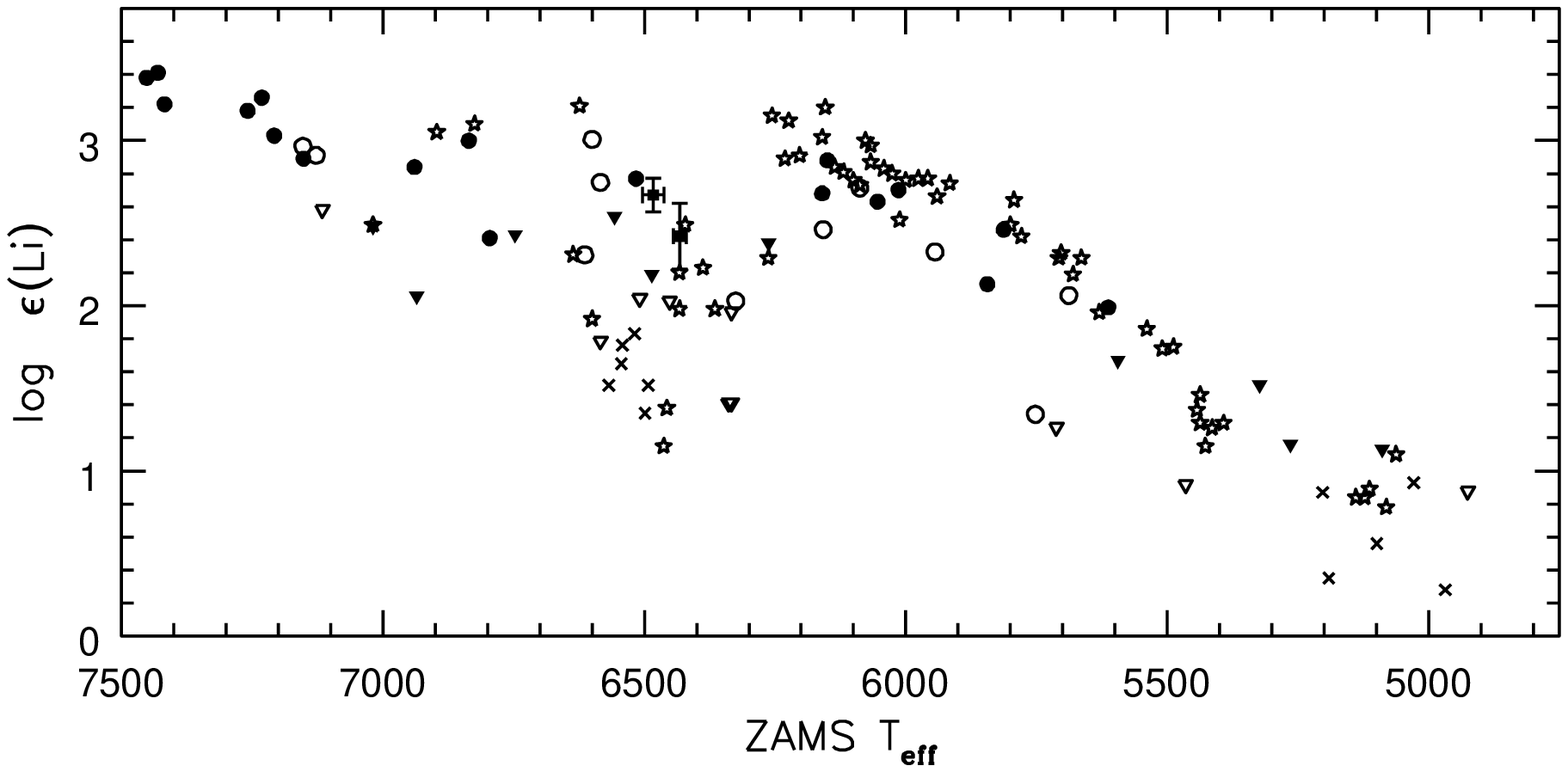}
\vskip -4.5cm
\caption[Li Dip plots]
{A(Li) versus ZAMS \teff\ plane containing: {\ }the V505 Per components 
(solid squares); objects in the 1.45Gyr-old \citep{AT2009} cluster \ngcseven\ 
(open circles; inverted open triangles are upper limits) taken from the reanalysis 
of \citet{balachandranp}; the 1.75Gyr-old \citep{AT2009} cluster 
\ngcthree\  (solid circles are detections; solid inverted triangles are upper 
limits) taken from \citet{AT2009}; and the 650 Myr-old Hyades cluster data (open 
stars are detections, crosses are upper limits) from \citet{balachandranp}.}
\label{fig:fig3}
\end{figure}

\end{document}